\documentclass[doublecol]{epl2}
\input epsf.sty
\usepackage{amsmath}
\usepackage{graphicx}

\title{Convection cells induced by spontaneous symmetry breaking}
\shorttitle{Convection cells induced by spontaneous symmetry breaking}
\author{M. Pleimling \and B. Schmittmann \and R.K.P. Zia}
\shortauthor{M. Pleimling \etal}
\institute{
\inst{1} Department of Physics, Virginia Tech, Blacksburg, VA 24061-0435 USA}

\abstract{
Ubiquitous in nature, convection cells are a clear signature of systems
out-of-equilibrium. Typically, they are driven by external forces, like
gravity (in combination with temperature gradients) or shear. In this
article, we show the existence of such cells in possibly the simplest
system, one that involves only a temperature gradient. In particular, we
consider an Ising lattice gas on a square lattice, in contact with two
thermal reservoirs, one at infinite temperature and another at $T$. When
this system settles into a {\em non-equilibrium} stationary state, many
interesting phenomena exist. One of these is the emergence of convection
cells, driven by spontaneous symmetry breaking when $T$ is set below the
critical temperature.}

\begin{document}
\maketitle

\section{Introduction}

In nature, convection cells can be found everywhere, existing at length
scales from the microscopic to the global. Rayleigh-Benard cells \cite{RBc}
are likely the most well known examples. Another are the vortices (in, e.g.,
cloud formations) created by the Kelvin-Helmholtz instability \cite
{KHi1,KHi2}. In all cases, external forces drive convection -- temperature
gradients and gravity in the former case and shear in the latter. Clearly,
none of these cells can persist under conditions of thermal equilibrium.
Instead, they are distinct signatures of systems far from equilibrium.
Motivated by our interest in understanding fundamental issues in
non-equilibrium statistical mechanics, we ask here: What are the minimal
conditions for convection cells to persist? The answer may provide valuable
insight into the essential ingredients associated with physics far from
equilibrium. To identify these conditions, we construct minimal systems,
seeking the equivalent of the Lenz-Ising model \cite{Ising} in the study of
phase transitions. In this spirit, we focus specifically on non-equilibrium 
\emph{steady states} (NESS), in which convection cells persist in a \emph{%
time-independent} manner. Here, we present possibly the simplest model with
such properties. Unlike the examples above, our system is taken out of
equilibrium by a temperature gradient \emph{alone}. In particular, we couple
two (spatial)\ sectors of an Ising lattice gas to two different thermal
baths, evolving with familiar dynamics and rates. Thus, the underlying
dynamics fully respect the Ising symmetry. With no other external forces,
our cells are the consequence of spontaneous symmetry breaking, i.e., phase
segregation.

The properties of the Lenz-Ising model, in its lattice gas version \cite
{Lee-Yang1,Lee-Yang2} and in thermal equilibrium, are well established.
Especially prominent is a second order phase transition at the Onsager \cite
{Onsager1, Onsager2} temperature, $T_O\cong 0.5673\,J/k_B$, in half-filled
systems in two dimensions (2D). To drive this model into a NESS and explore
its behavior far from equilibrium, many authors have introduced a variety of
changes to its usual dynamics, e.g., allowing a uniform bias along one of
the lattice directions \cite{KLS1,KLS2} or using two different $T$'s for
exchanges along the two axes \cite{KETT1,KETT2,KETT3,KETT4}. A wide range of
novel phenomena emerged, leading to a better understanding of NESS in
general \cite{DL17a,DL17b}. Our approach here follows similar lines, in that
we couple the system to two thermal baths, but in a manner that is far more
common in nature. As in earlier studies of NESS, the results are novel and
surprising, with the most notable being the emergence of convection cells.
In the remainder of this letter, we define our model, present simulation
results and quantitative measures for these cells. Next, we provide an
exactly solvable (small) system which displays such cells, as well as steps
toward a field theoretic framework for analyzing the phenomenon. Deferring
systematic details to a future publication, we close with a discussion and
an outlook for future studies.

\section{Model specifications}

Our system consists of a 2D Ising model with the usual nearest-neighbor (NN)
interaction on a $L_x\times L_y$ square lattice (with $L_x$ and $L_y$ both
even, for simplicity). In a lattice gas, a site $\left( x,y\right) $ may be
occupied by a particle or left empty, i.e., $n\left( x,y\right) =1$ or $0$.
The interparticle interactions are attractive, i.e., $\mathcal{H}=-J\sum
n\left( x,y\right) n\left( x^{\prime },y^{\prime }\right) $ with the sum
over NN sites and $J>0$. The system evolves with ordinary Kawasaki dynamics 
\cite{Kawasaki}: A NN pair (also referred to as a ``bond'') is chosen at
random and exchanged according to the familiar Metropolis rates: $\min
[1,e^{-\Delta \mathcal{H}/k_BT}]$, where $\Delta \mathcal{H}$ is the change
in $\mathcal{H}$ due to the exchange. Specifying the boundary conditions and
starting with an initial configuration, this system will settle into thermal
equilibrium, with well-known Ising lattice gas properties. In particular,
for $T<T_O$, a half-filled system displays co-existence of phases, i.e., a
high-density domain separated from a low-density one by well defined
interfaces. With periodic boundary conditions (PBC), each domain is a strip,
generally aligned along an axis so that its interfaces span the smaller of $%
L_x,L_y$.

To impose coupling to \emph{two} thermal baths, we partition the lattice
into two sectors (say, the ``left'' and ``right'' half-planes: $x\leq L_x/2$
and $>L_x/2$) and use different $T$'s for each.\footnote{
Other non-equilibrium Ising models coupled to two temperature baths, 
especially using Glauber spin flip dynamics, have also been studied. 
For a review and references, see e.g., Section VII-B in \cite{DL17a}.}
This form of drive, with a
macroscopic temperature gradient, is obviously ubiquitous, from stove-top
cooking to large scale weather patterns. Specifically, we set the right bath
temperature, $T^{\prime }$, to infinity (for simplicity), but leave the left
bath at a controllable $T$. In other words, if a particle-hole pair lying 
\emph{strictly} in the left sector is chosen, they are updated precisely as
in the ordinary Ising case. Otherwise, the pair is always exchanged ($%
e^{-\Delta \mathcal{H}/\infty }=1$). A pair straddling the sectors, $\left(
L_x/2,y\right) $-$\left( 1+L_x/2,y\right) $, is also exchanged (again, for
simplicity). As for the boundary conditions, the simplest would be periodic
(PBC). However, to facilitate the measurement of convection cells, we impose
pinned conditions at the $x$-boundaries. Specifically, we introduce an extra
column of \emph{immobile} objects at $x=0$: particles at the ``bottom'' ($%
y\in \left[ 1,L_y/2\right] $) and holes at the ''top'' ($y\in \left[
1+L_y/2,L_y\right] $). The $y$-boundaries are periodic. Below, we will
discuss how to detect these cells in a system with full PBC. Here, let us
emphasize that the pin introduced here plays an identical role for an Ising
model in equilibrium, in which the average magentisation of a finite system
with full PBC is \emph{zero }for all $T$.   

In our simulations, a Monte Carlo Step (MCS) consists of $L_xL_y$ attempts
to exchange pairs. Starting with a random, half-filled configuration, we
discard $1.6\times 10^6$ MCS so that the system has relaxed into a steady
state. The next $K$ MCS are designated as our ``run''. Mostly, we have $%
K=3\times 10^7$ and carry out 80 runs for better statistics. In addition to
the usual measurements of the average density $\langle n\left( x,y\right)
\rangle $, we record \emph{every successful exchange} over the entire run.
The latter allows us to define \emph{net currents} across each bond, e.g., $%
j_x\left( x,y\right) $ is the total number of traverses by a particle from $%
\left( x,y\right) $ to $\left( x+1,y\right) $, minus such traverses by a
hole, divided by $K$. As a check, for each site we compute \emph{div }$\vec{j%
}\equiv j_x\left( x,y\right) -j_x\left( x-1,y\right) +j_y\left( x,y\right)
-j_y\left( x,y-1\right) $ and see that it is precisely $\left[ n\left(
x,y;K\right) -n\left( x,y;0\right) \right] /K$. Thus, it vanishes in the $%
K\rightarrow \infty $ limit.

By contrast, there is no such constraint on the vorticity $\omega \equiv $ 
\emph{curl }$\vec{j}$ (a scalar in 2D). On the lattice, $\omega \left(
x,y\right) $ can be associated with the plaquette centered at $\left(
x+\frac 12,y+\frac 12\right) $. A discrete version of $\oint \vec{j}\cdot d%
\vec{\ell}$ around the square, 
\begin{equation}
\omega \left( x,y\right) \equiv j_x\left( x,y\right) +j_y\left( x+1,y\right)
-j_x\left( x,y+1\right) -j_y\left( x,y\right)
\end{equation}
is positive for counter-clockwise circulation. For an equilibrium system,
the average current, $\langle \vec{j}\rangle $, must vanish, and so, $%
\langle \omega \rangle =0$. Being stochastic, $\omega \left( x,y;t\right) $
should be a random walk in time. Thus, with our definition of $\vec{j}$, $%
\langle \omega ^2\rangle $ is expected to vanish as $1/K$. For a system in a
NESS, however, probability currents are non-trivial in general \cite
{ZS07a,ZS07b}, so that both $\langle \vec{j}\rangle $ and $\langle \omega
\rangle $ may also be non-trivial. Of course, in our model, we do not expect
the presence of uniform, \emph{global} currents or vorticity 
(such as in the driven
lattice gas \cite{KLS1,KLS2}). Instead, the symmetries dictate the presence
of convection \emph{cells} in equal and opposite pairs.

\section{Simulation results}

Focusing on the existence of $\omega $, we defer our systematic study to a
future publication and present only one set of results here, namely, $T\cong
0.88T_O$ with $L_x=20,50,100$ and $L_y=20,50,100,200,400$. Fig. 1a shows a
typical configuration in a $100\times 100$ system and Fig. 1b shows a plot
of $\langle n\left( x,y\right) \rangle $. Not surprisingly, we find phase
segregation in the left sector, with a high-density domain ``at the bottom''
(due to the pinned BC). Also expected is a homogeneous, half-filled state on
the right. Across the sectors, there are density gradients which tend to
drive particles/holes from the high/low density domain on the left to the
right sector. This leads to emerging local excesses in the right sector
which, in turn, are also unsustainable, due to the underlying dynamics, so
that we expect nontrivial $\vec{j}$, $\omega $, and persistent convection
cells. Such a scenario is sketched in Fig 1c.

\begin{figure}[t]
\centerline{\epsfxsize=3.60in\ \epsfbox{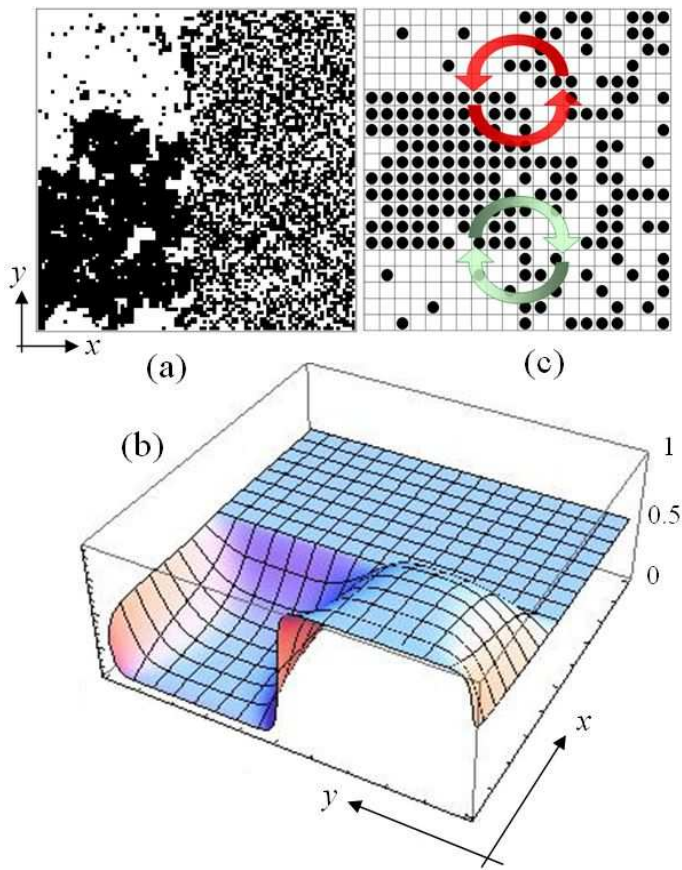}}
\caption{ (a) Typical configuration of a $100\times 100$ system, with $%
T\cong 0.88T_O$ and $T^{\prime }=\infty $. (b) The average density profile.
(c) A sketch of a $20\times 20$ case, illustrating the presence of the
convection cells. For clarity, this system is displaced ``upwards'' by $%
L_y/4 $. Thus, the top row shown here is the $y=15$ line. }
\label{fig1}
\end{figure}

To quantify $\vec{j}$ and $\omega $, we first verify that the net number of
exchanges across each bond indeed increases linearly with $K$ for each run.
Averaging over the 80 runs, we establish a non-vanishing $\langle \vec{j}%
\left( x,y\right) \rangle $. To display a vector field is typically
cumbersome. However, in 2D, a convenient quantity is the \emph{scalar }%
stream function $\psi \left( x,y\right) $, the \emph{curl} of which is $\vec{%
j}$. In Fig. 2a, we show $\psi $ for a $50\times 50$ system, illustrating
that the currents are highly \emph{nonlocal}, i.e., spread out over the
entire lattice\footnote{In order to appreciate this remarkable observation,
let us emphasize that the currents would be identically zero everywhere, had the 
two temperatures been the same. Our dynamics violates detailed balance at a local level 
(three columns of exchanges), yet the currents - a prominent signature of a non-equilibrium system - are nonlocal.}!
For easier interpretation, we note that $j_x=\partial _y\psi 
$ and $j_y=-\partial _x\psi $. Hence, $\vec{j}$ can be visualized from $\psi 
$ by taking its gradient and rotating the resultant vectors by $90^{\circ }$%
, so that the peak at $x=y=25$ corresponds to the center of the $\omega >0$
vortex (red online) in Fig 1c. Another perspective is provided by the
associated $\omega $, shown in Fig 2b. Notably, it is essentially localized
at the sector boundary ($x=25$), with peaks/dips centered on the convection
cells. Within each sector, just one or two columns away from the boundary, $%
\omega $ drops precipitously to noisy backgrounds. To highlight this
localization and its secondary characteristics, we illustrate $\omega $ in a 
$20\times 400$ system (sector boundary at $x=10$). In Fig. 2c, we plot $%
\omega \left( x,y\right) $ against $y$ for several values of $x$. Note that
(i) the width of the primary vortex in $y$ is only about $10$, (ii) the
width in $x$ is $O\left( 1\right) $, and (iii) there are two secondary,
counterflow cells here.

\begin{figure}[t]
\centerline{\epsfxsize=3.60in\ \epsfbox{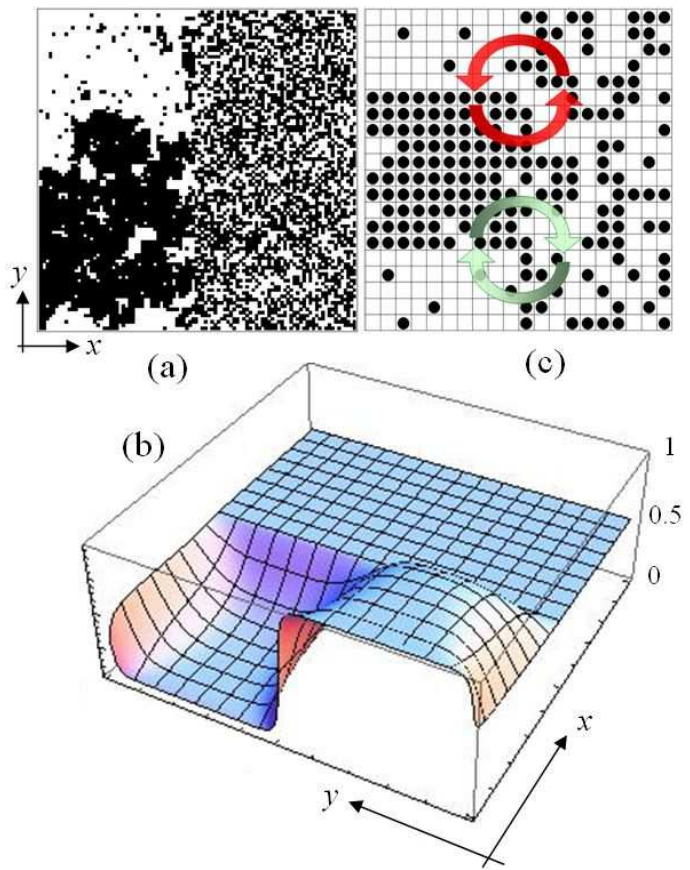}}
\caption{ Persistent currents and vortices present in a $50\times 50$
system, with $T\cong 0.88T_O$ and $T^{\prime }=\infty $, characterized by
(a) the stream function $\psi $ and (b) the vorticity $\omega $. (c)
Vorticity in a $20\times 400$ system: $\omega \left( x,y\right) $ \emph{vs.} 
$y$ for several values of $x$, showing that it is essentially localized at
the sector boundary ($x=10$), with counter-flow cells about $10$ lattice
spacings on either side of the primary vortex. Arbitrary units are used for
both $\psi $ and $\omega .$ }
\label{fig2}
\end{figure}

The contrast between $\psi $ and $\omega $ should caution us on how best to
describe ``non-equilibrium effects'' -- some aspects being system-wide and
others appearing to be localized. Of course, such constrasts are well
known: In electrostatics, the potential of a localized charge ranges
throughout space. Indeed, the analogy here is precise, since $\nabla ^2\psi
=-\omega $. To emphasize the absence of vortices in an equilibrium system,
we consider a model which is \emph{superficially} similar to ours, by
setting $J=0$ on all bonds centered on $x\geq \frac{L_x+1}2$, but updating
with a single $T$ everywhere. The dynamical rules \emph{within} the right
sector here are identical to our model above. The only difference between
the two lies with the updates on the three sets of bonds centered on $x=%
\frac{L_x}2,\frac{L_x\pm 1}2$. Simulations (to be reported elsewhere) indeed
show all $\omega $'s distributed around zero, with a width vanishing as $%
K^{-1/2}$.

\section{Theoretical considerations}

To formulate a microscopic theory for our model is facile: A master equation
can be written in a line or two. However, to find the stationary
distribution is far from trivial. Nevertheless, we can offer some insight by
considering the exact solution for the simplest possible case, namely, a $%
2\times 2$ lattice with just $2$ particles. Now, for just two rows, we
impose open BC in $y$ (since PBC would be meaningless) and so, just one bond
is relevant, for coupling and exchange across the rows. With pin BC in $x$,
particle-hole exchanges can take place on only $4$ bonds. Thus, their
associated $\langle \vec{j}\rangle $'s in the steady state must share the
same magnitude and only a single $\omega $ exists. With just 6
configurations ($i=1,...6$), it is straightforward to write the master
equation and find the stationary probabilities exactly. From here, we can
compute the frequency of any exchange to take place and arrive at $\langle
\omega \rangle =2\left( 1-e^{-4J/k_{B}T}\right) /\left(
7+5e^{-4J/k_{B}T}\right) $. To appreciate the significance of having two
baths, we repeat this computation with $T^{\prime }<\infty $ and find that
the first term in the numerator is $e^{-4J/k_BT^{\prime }}$. Hence, $\langle
\omega \rangle $ vanishes when $T^{\prime }=T$, as expected.

A concern may be raised that, together with the open BC, the pinned column
affects particles much like gravity and so, our convection cell is no
different from those observed daily in the atmosphere. Indeed, for the $%
2\times 2$ system, the pin, along with the extremely small size, has serious
consequences, such as the vortex persisting for \emph{all} $T<\infty $. To
untangle the effects of pin \emph{vs}. size, we turn to slightly larger
systems, of size $2\times \ell $ with $\ell \leq 6$, and \emph{open} BC at
both (left and right) edges. Now, we can re-impose PBC in $y$ so that
translational invariance is restored. Of course, this delocalizes the vortex
and the \emph{average} vorticity, $\langle \omega \rangle $, is necessarily
zero. Nevertheless, despite the absence of a gravity-like pin, convection
cells are still present and can be detected through a density-vorticity
correlation, such as $\langle n\omega \rangle $. Using exact numerical
techniques \cite{Sid}, non-trivial values for $\langle n\left( 1,y\right)
\omega \left( 1,y\right) \rangle $ and $\langle n\left( 1,y\right) \omega
\left( 1,y+1\right) \rangle $ emerge, persisting again for \emph{all} finite 
$T$. We believe that this behavior has a counterpart in an equilibrium Ising
model, e.g., in the two-spin correlation $\langle s\left( x,y\right) s\left(
x+\frac{L_{x}}{2},y+\frac{L_{y}}{2}\right) \rangle $. This quantity is
non-zero for \emph{all} $T<\infty $ for a \emph{finite} system. The
difference between the two sectors $T<T_{O}$ and $T\geq T_{O}$ emerges only
for $L\rightarrow \infty $: the correlation vanishes in the latter and
remains non-zero for $T<T_{O}$. Of course, a systematic finite-size analysis
of $\langle n\omega \rangle $ will address these issues decisively.
Simulation studies for this purpose are in progress.

Small systems, though exactly solvable, are clearly of limited value for
understanding large scale collective behavior. For the latter, a standard
theoretical approach is to exploit Langevin equations for a mesoscopic
continuum particle density $\rho \left( \vec{x},t\right) $. For an Ising
lattice gas evolving towards thermal equilibrium, a well established route
is model B \cite{HHM}: $\partial _t\rho \propto \nabla ^2\left( \delta 
\mathcal{F}/\delta \rho \right) +\eta $, where $\mathcal{F}$ is the
Landau-Ginzburg free energy functional and $\eta $ is a conserved noise. In
this approach, the vorticity \emph{never }appears, since the right hand side
is just the divergence of the current. Although \emph{curl} $\vec{j}$ can
never affect the evolution of $\rho $, it can be significant, especially for
non-equilibrium systems. However, if we take the form for $\vec{j}$ from
model B above, then (the deterministic part of) \emph{curl} $\vec{j}$
vanishes identically and cannot be captured. Instead, it is necessary to
include the mobility factor, $\sigma $, in the functional form of $\vec{j}:$ 
\begin{equation}
\vec{j}\left[ \rho \right] =\sigma \left[ \rho \right] \left\{ -\vec{\nabla}%
\frac{\delta \mathcal{F}}{\delta \rho }\right\} \,\,.
\end{equation}
Such a factor has been exploited in other studies, especially in systems
with external drives like gravitational or electric fields \cite
{sigma1,sigma2}. Typically, $\sigma $ is assumed to be $\rho \left( 1-\rho
\right) $ for the lattice gas \cite{DL17a,DL17b}. To have \emph{curl} $\vec{j%
}$ $\neq 0$, a further ingredient is necessary, namely, explicit spatial
dependence in either $\sigma $ or $\mathcal{F}$, or both. Otherwise, \emph{%
curl} $\vec{j}$ will be proportional to $\vec{\nabla}\rho \times \vec{\nabla}%
\rho \equiv 0$. In our case, we would model the two sectors coupled to baths
with $T\neq T^{\prime } $ by having an explicit $\vec{x}$-dependence in $%
\mathcal{F}\left[ \rho ,\vec{x}\right] $. (It is also simple to consider
models with $\sigma \left[ \rho ,\vec{x}\right] $, which would induce shear
when driven.) One advantage of this approach is that we expect $\vec{\nabla}%
T\left( x\right) $ to appear, so that \emph{curl} $\vec{j}$ will be
localized to the sector boundary (for the model presented here). 
Work is in progress to turn this intuitive
picture into a developed mesoscopic theory.

\section{Summary and Outlook}

In this letter, we report the presence of convection cells in a minimal
model: an Ising lattice gas coupled to two thermal baths. Unlike cells
common in nature, they are not driven by external forces such as gravity or
shear. Instead, the microscopic dynamics obeys the full Ising (up-down, or
particle-hole) symmetry, so that the convection cells here are induced by
the spontaneous breaking of this symmetry (and the ensuing density
gradients). Through simulations and an exact analysis of small systems, we
investigate a natural quantity to describe convection cells -- the
vorticity, $\omega $, or \emph{curl} of the current density. To our
knowledge, $\omega $ has never been studied previously in systems where
particles are not associated with momenta. We believe our findings offer a
new perspective on the vorticity.

A range of interesting questions natural arise from this study, some of
which can be explored readily. Are the vortices affected by the details of
the exchanges across the sector boundary? How does the spread of $\omega $
along the boundary ($O\left( 10\right) $ in Fig 2c) depend on $L_x$? Are the
properties of the interface (between high- and low-density domains on the
left) similar to those in equilibrium, or are they drastically different?
Are there any novel critical properties associated with $\omega $, which
vanishes along with the spontaneous symmetry breaking at high $T$? What is
the best characterization of the energy flux from the right to the left
sector? Then there are, of course, deeper issues which deserve careful
thought and serious effort. What is the origin and behavior of the secondary
counterflows? And is there a series of smaller and smaller vortices? How do
the phenomena found here behave in the thermodynamic limit? Can the density
profiles (e.g., in Fig 1c) be quantitatively understood?

Apart from vorticity and convection cells, our two-temperature model
displays many other intriguing and counterintuitive properties. In
particular, we found an extremely surprising phenomenon in a fully periodic, 
$T>T_O$ system, in which only a strip of a few columns is coupled to $%
T^{\prime }=\infty $, namely, the particle density within the strip being a 
\emph{bi-modal distribution} (symmetric around 1/2)! So far, this remarkable
behavior is yet to be understood. From such surprises and the presence of
convection cells, we can expect this deceptively simple two-temperature
lattice gas to present us with further puzzling challenges. We believe that
solving these puzzles will provide critical clues to formulating an
overarching theoretical framework for non-equilibrium statistical mechanics.

\acknowledgments
We thank Henk Hilhorst and Zoltan Toroczkai for illuminating discussions.
This research is supported in part by grants from the US National Science
Foundation, DMR-0705152 and DMR-0904999.

\end{document}